\documentstyle[sprocl]{article}

\def\Journal#1#2#3#4{{#1} {\bf #2}, #3 (#4)}

\def\NPB{{\em Nucl. Phys.} B}
\def\PLB{{\em Phys. Lett.}  B}
\def\PLA{{\em Phys. Lett.}  A}
\def\PRL{\em Phys. Rev. Lett.}
\def\PRT{\em Phys. Rept.}
\def\PRD{{\em Phys. Rev.} D}

\def\be{\begin{equation}}
\def\ee{\end{equation}}
\def\bea{\begin{eqnarray}}
\def\eea{\end{eqnarray}}

\def\etal{{\it et al,\/}\ }

\begin{document}

\rightline{OUTP--00--11P}
\rightline{TPI--MINN--00/12}
\rightline{UMN--HEP--1846}
\rightline{hep-ph/0003156}
\rightline{March 2000}
\title{Duality, Equivalence, Mass\\and\\ The Quest For The Vacuum
\footnote{Invited talk presented at PASCOS 99, Lake Tahoe, CA December
10--16 1999.}
}
\author{Alon E. Faraggi}

\address{Department of physics, University of Minnesota,
Minneapolis, MN 55455, USA\\and\\
Theoretical physics, Oxford University, 1 Keble Road,
Oxford OX1 3NP, UK\\E-mail: faraggi@mnhepo.hep.umn.edu}


\maketitle\abstracts{
I contemplate the possibility that the mismatch
between the maximally symmetric point (the free fermionic point)
and the strictly self--dual point in the Narain moduli space
plays a role in the string vacuum selection. The
role of self--duality in the recent formulation
of quantum mechanics from an equivalence postulate,
and the new perspective that it offers on the foundations
of quantum gravity and the origin of mass, are discussed.}

The central issue in elementary particle physics
is the nature of the electroweak symmetry breaking mechanism.
Perhaps not unrelated, and of equal importance,
is the formulation of the consistent synthesis of
gravity and quantum mechanics from fundamental
physical postulates.
String theory and its non--perturbative generalizations
constitute the most advanced such attempts.

The Standard Model, and many of its contemporary
theoretical extensions, utilize fundamental scalar
representations to break the electroweak symmetry.
To my knowledge all of the existing string theories
give rise to fundamental scalar representations.
Therefore, as a matter of classification we may
classify all the string theories as theories
that include fundamental scalars. The precise
realization of the scalar state in nature,
and its role in electroweak symmetry breaking
is, however, at present an experimentally unresolved issue.

The experimental success of the Standard Model raises the problem
of understanding the origin of its structure and parameters.
The Standard Model multiplets suggest the embedding in
Grand Unified Theories. The flavor sector does
not have an appealing explanation in this framework, and
necessitates further ad--hoc assumptions. A more
constraining framework is sought in the context
of theories which unify gravity with the gauge interactions.
Such a concrete framework is given by string theories.
In this conference the talks by Fernando Quevedo and
Jerry Cleaver report on some current efforts.
Since my work in this area is also covered in Cleaver's
talk, I only discuss here briefly some of the interesting aspects.

String theories give rise to a huge number of potentially
viable vacua. Selecting the correct one among them is
a daunting task. One may further question whether
in the lack of complete understanding of the theory
such an endeavor is not futile to begin with.
However, with present day understandings a reasonable
goal is to use the low energy data to single out
the string theories that most closely resemble the real
world. The important guide in this quest is
the multiplet structure of the Standard Model.
It is natural to seek superstring models
which preserve the $SO(10)$ embedding of the Standard
Model spectrum. The $SO(10)$ symmetry however can be broken directly
at the string theory level rather than at the level of the
effective field theory. Additionally we must
impose the existence of three chiral generations.

The heterotic--string models constructed in the free fermionic
formulation naturally achieve both of these criteria \cite{rffm}.
Furthermore, a generic consequence of perturbative string models is the
existence of numerous massless states beyond the spectrum of the Minimal
Supersymmetric Standard Model. Many
of these states carry fractional electric charge
and consequently must be decoupled from the low
energy spectrum. Recently, it was demonstrated,
in the FNY heterotic string model \cite{fny},
that free fermionic models also give
rise to models in which all the states beyond the
MSSM decouple from the low energy spectrum at
or slightly below the string scale \cite{cfn}. More on
this is discussed in Cleaver's talk.

It should be emphasized that the success of the FNY
model in producing a Minimal Standard Heterotic String Model
should not be viewed as implying that the FNY model
is the correct string vacuum. Indeed, it is preposterous
at present to suggest that any three generation string model
is the true string vacuum. The free fermionic models, however,
give rise to a large class of three generation models.
Therefore, it makes sense to extract the features that
underly this large class of models. While it is
preposterous to suggest that any string model is the true
string vacuum, it is not unplausible that the true string
vacuum shares some of the properties of the realistic free
fermionic models. It is these properties that we would like to
extract. It is also quite plausible that these underlying
properties may also offer a clue to the dynamical mechanism
which selects the string vacuum. It is important to remark
that the eventual true string vacuum need not necessarily
be an heterotic string. Indeed, it may not even be a string
at all! However, the heterotic string may still provide
a useful probe to the vital properties of the true vacuum.

The free fermionic models are built by specifying a set
of boundary condition basis vectors for all world--sheet
fermions and the one--loop GSO projection coefficients \cite{FFF}.
The NAHE set, $\{{\bf1},S,b_1,b_2,b_3\}$,
is a set of boundary condition basis vectors which is common in all the
realistic free fermionic models.
The important aspect of the NAHE set is its correspondence
with $Z_2\times Z_2$ orbifold compactification \cite{ztwo}. However,
this $Z_2\times Z_2$ orbifold act at a very special point
in the Narain moduli space! At this point the symmetry
which arises from the six dimensional compactified lattice
is maximized! That is, at the point in the Narain moduli
space where the internal compactified dimensions can be
represented as free fermions propagating on the string
world--sheet one obtains the maximal symmetry,
which for $T^6$ is $SO(12)$.

The free fermionic point in the Narain moduli space is a maximally
symmetric point. However, as is well known toroidal
compactifications of string theories possess a duality
symmetry under the interchange of winding and momentum
modes and the generalization of $R\leftrightarrow {1/R}$
duality \cite{giveon}. The free fermionic point is realized
for a specific value of $R$. Thus, the maximally symmetric
point is realized at a specific value of $R$ in the moduli
space. Under the duality interchange
there exist a value for $R$ which is self--dual.
The strictly self--dual point is realized at the point with
$(G+B)^2=I$, where $G$ and $B$ are the metric and antisymmetric
tensor, respectively. The enhanced symmetry at the strictly
self--dual point of a six dimensional compactified torus is
$SU(2)^6$. We see that there is a mismatch between
the most symmetric point, where the internal dimensions
are realized as free fermions on the world--sheet,
and the strictly self--dual point.

We can contemplate that the mismatch between the maximally
symmetric point and the strictly self--dual point plays a
role in the dynamical mechanism which selects the string vacuum.
One possibility (which I am not sure is correct, but can be
examined explicitly by studying the relevant partition functions)
is that the effect of the orbifold twisting is
to move the free fermionic point to the self--dual point.
It is interesting to note that after the $Z_2\times Z_2$
orbifold twisting the $SO(12)$ symmetry is broken to $SU(2)^6$,
which is the enhanced symmetry at the strictly self--dual point.
This scenario would then suggest a dynamical reason why the
$Z_2\times Z_2$ orbifold is selected.
There are however several caveats to this proposal.
The first is that supersymmetry is unbroken.
We may envision the possibility
that supersymmetry is broken dynamically by hidden sector
condensation rather than by the mechanism which selects
the string vacuum. In this regard
it would also be of interest to
study the properties of self--dual string models that are not
supersymmetric. A second caveat is that
the argument above is in the framework of perturbative
string compactifications. However, from string
dualities we know that there is a nonperturbative
structure which underlies the different string theories.
Moreover, we know that in this structure an eleventh
dimension plays a key role. We may envision
the possibility that the argument holds also for the
eleventh dimension. This would seem to suggest
why the Horava--Witten theory \cite{hw} is the viable framework.
However, it would also suggest that string coupling
is of order one, which seems to be in contradiction
with the coupling extracted from extrapolation
of the gauge couplings from low energies which
yield a smaller value. The possible resolution may
be the existence of additional vector--like matter
states, beyond the MSSM, in the desert \cite{dienes}.

Duality and self--duality also play a key role in the recent
formulation of quantum mechanics from an equivalence
postulate \cite{fm1,flyod,vancea}.
This is seemingly unrelated to the string
program. However, I suggest that this is not the case.
As expounded above the central issues of particle physics
are the problem of mass and the formulation of quantum gravity
from fundamental postulates. Although string theory provides a
useful probe for quantum gravity, surely it does not yet
provide such a satisfactory formulation, even with the
deeper understandings gained from string dualities.
Moreover, at the basic observational level none of the
current approaches to quantum gravity provides a compelling
solution to the vacuum energy problem. It seems to me that
all of the current approaches entail, in one form or another,
a careful bookkeeping of the energy checkbook, and therefore
in the end amount to some form of fine tuning. However,
rather than a slight adjustment what may be needed is
a new Copernican revolution, which in our case would be a new
view of the Hilbert space. I propose that the formulation
of quantum gravity from the equivalence postulate offers
such a new view. A key question in this respect is,
how do the basic particle properties arise in this formalism.

An important facet of the equivalence postulate derivation is the
phase--space duality, which is manifested in this formalism due to the
involutive nature of the Legendre transformation. The phase--space
duality arises due to the defining relation between the
dual variables, $p=\partial_q{\cal S}_0$, through the generating
function ${\cal S}_0$. However, the fact that the Legendre
transformation is not defined for linear functions, {\it i.e.} for
physical systems with ${\cal S}_0=A q+B$, implies that the
Legendre duality fails for the free system and for the
free system with vanishing energy. The Legendre phase--space
duality and its breakdown for the free system are intimately
related to the equivalence postulate, which states
that all physical systems labeled by the function
${\cal W}(q)=V(q)-E$, can be connected by a coordinate
transformation, $q^a\rightarrow q^b=q^b(q^a)$, defined
by ${\cal S}_0^b(q^b)={\cal S}_0^a(q^a)$.
This postulate implies that there
always exist a coordinate transformation connecting
any state to the state ${\cal W}^0(q^0)=0$. Inversely, this means
that any physical state can be reached from the
state ${\cal W}^0(q^0)$ by a coordinate transformation.
This postulate cannot be consistent
with classical mechanics. The reason being that in Classical
Mechanics (CM) the state ${\cal W}^0(q^0)\equiv0$ remains a fixed
point under coordinate transformations. Thus, in CM it
is not possible to generate all states by a coordinate
transformation from the trivial state. Consistency of the
equivalence postulate implies the modification of CM,
which is analyzed by a adding a still unknown function
$Q$ to the Classical Hamilton--Jacobi Equation (CHJE).
Consistency of the equivalence postulate fixes the
transformation properties for ${\cal W}(q)$,
$$
{\cal W}^v(q^v)=
 \left(\partial_{q^v}q^a\right)^2{\cal W}^a(q^a)+(q^a;q^v),
$$
and for $Q(q)$,
$$
 Q^v(q^v)=\left(\partial_{q^v}q^a\right)^2Q^a(q^a)-(q^a;q^v),
$$
which fixes the cocycle condition for the inhomogeneous term
$$
(q^a;q^c)=\left(\partial_{q^c}q^b\right)^2[(q^a;q^b)-(q^c;q^b)].
$$
The cocycle condition is invariant under M\"obius transformations
and fixes the functional form of the inhomogeneous term.
The cocycle condition is generalizable to higher, Euclidean or
Minkowski, dimensions,
where the Jacobian of the coordinate transformation extends
to the ratio of momenta in the transformed and original systems.

The identity
$$
({\partial_q{\cal S}_0})^2=
\hbar^2/2\left(\{\exp(i2{\cal S}_0/\hbar,q)\}-\{{\cal S}_0,q\}\right)
$$
is the second key ingredient in the
equivalence postulate formulation.
Making the identification
$$
{\cal W}(q)= V(q) - E = -{\hbar^2/{4m}}\{{\rm e}^{(i2{\cal S}_0/\hbar)},q\},
$$
and
$$
{Q}(q)=  {\hbar^2/{4m}}\{{\cal S}_0,q\},
$$
we have that
${\cal S}_0$ is solution of the Quantum Stationary
Hamilton--Jacobi Equation (QSHJE),
$$
({1/{2m}})\left({{\partial_q S}_0}\right)^2+
V(q)-E+({\hbar^2/{4m}})\{{\cal S}_0,q\}=0,
$$
where $\{,\}$ denotes the Schwarzian derivative.
{}From the identity we deduce that the trivializing
map is given by $q\rightarrow {\tilde q}=\psi^D/\psi$,
where $\psi^D$ and $\psi$ are the two linearly independent
solutions of the corresponding Schr\"odinger equation \cite{fm1}.
We see that the consistency of the equivalence postulate
forces the appearance of $\hbar$ as a covariantizing parameter.

The remarkable property of the QSHJE, which distinguishes
it from the classical case, is that it admits non--trivial solution
also for the trivial state, ${\cal W}(q)\equiv0$.
In fact the QSHJE implies that ${\cal S}_0=constant$ is
not an allowed solution. The fundamental characteristic
of quantum mechanics in this approach is that ${\cal S}_0\ne Aq+B$.
Rather, the solution for the ground state, with $V(q)=0$ and $E=0$,
is given by
$$
{\cal S}_0=i\hbar/2\ln q,
$$
up to M\"obius transformations. Remarkably, this quantum
ground state solution coincides with the self--dual state
of the Legendre phase--space transformation and its dual.
Thus, we have that the quantum self--dual state plays a pivotal
role in ensuring both the consistency of the equivalence
postulate and definability of the Legendre phase--space
duality for all physical states. The association of the
self--dual state and the physical state with $V(q)=0$ and
$E=0$ provides a hint that the equivalence postulate
and Legendre phase--space duality may shed new light
on the nature of the vacuum.

A second remarkable consequence of the equivalence postulate
is that it implies energy quantization for bound states
without assuming the probability interpretation of the
wave--function. Consistency of the equivalence postulate
implies that the trivializing map, $q\rightarrow{\tilde q}=\psi^D/\psi$
should be continuous on the extended real line. It is then seen that
this condition is synonymous to the requirement that the physical
solution of the
corresponding Schr\"odinger equation admits a square integrable
solution, without assuming the probability interpretation of
the wave function. The equivalence postulate
formalism may therefore indeed offer an entire new
perspective on the origin of the Hilbert space structure.
The relation of the formalism to unifirmization theory and
Riemann surfaces  suggests that the Hilbert space structure
has a quantum--gravitational origin.

The equivalence postulate derivation may also shed light
on the quantum origin of mass. The generalization of
the Schwarzian identity to the relativistic case with
a vector potential is,
$$
\alpha^2(\partial{\cal S}-eA)^2={D^2(Re^{\alpha{\cal S}})/
(Re^{\alpha{\cal S}})}-{\Box R/ R}-({\alpha/ R^2})
\partial\cdot(R^2(\partial {\cal S}-eA)),
$$
where $\alpha=i/\hbar$, $D$ is a covariant derivative,
and $\partial\cdot(R^2(\partial{\cal S}-eA))=0$ is a
continuity condition. The $D^2(Re^{\alpha{\cal S}})/
(Re^{\alpha{\cal S}})$ term is associated with the
Klein--Gordon equation. In this case ${\cal W}(q)=1/2 m c^2$.
{}From the equivalence postulate it follows that masses
of elementary particles arise from the inhomogeneous
term in the transformation of the ${\cal W}^0(q^0)\equiv0$
state, {\it i.e.}
$$
1/2 m c^2= (q^0;q).
$$
{}From this perspective we may speculate that scalar particles
and symmetry breaking represent a particular realization
of the geometrical transformation $q^0\rightarrow q$.
Obviously, this interpretation offers new possibilities
to understand how particle properties are generated from
the vacuum. Generalizing the Schwarzian identity to
curved space will provide the equivalence postulate
approach to quantum gravity. Similarly, the identity
can be extended to include fermions. The more interesting
question, however, is to understand how the fermionic
degree of freedom, which has no classical counterpart,
arises from the consistency of the equivalence postulate.
We anticipate, once again, that the clue is given in
the identity itself.

\section*{Acknowledgments}

I would like to thank Gaetano Bertoldi, Jerry Cleaver,
Dimitri Nanopoulos, Joe Walker
and especially Marco Matone, for collaboration on part of the work
reported in this paper, and
Tonnis ter Veldhuis for comments on the manuscript.
This work is supported in part by DOE grant No. DE--FG--0287ER40328.

\end{document}